\documentclass[%
twocolumn,10pt,
 longbibliography,
 amsmath,amssymb, superscriptaddress,
 aps,
 pre,
]{revtex4}

\usepackage{float}
\usepackage{mathtools}
\usepackage{amsmath,amssymb}
\usepackage{graphics,graphicx,color, calc}
\usepackage{xcolor}
\usepackage{dcolumn}
\usepackage[english]{babel}
\usepackage{mathrsfs}
\usepackage[mathcal]{eucal}
\usepackage{bm}
\usepackage{lipsum}
\usepackage{xr}
\usepackage{stackengine}
\usepackage{booktabs}
\usepackage{soul}
\usepackage{xr-hyper}   
\usepackage{hyperref}
\usepackage[normalem]{ulem}

\DeclareMathOperator{\sgn}{sgn}

\begin{document}

\title{Effects of Non-reciprocity on Coupled Kuramoto Oscillators}

\author{Shaon Mandal Chakraborty}
\affiliation{Soft Condensed Matter Group, Raman Research Institute, Bengaluru 560080, Karnataka, India}

\author{Bibhut Sahoo}
\affiliation{Institute for Theoretical Physics, University of G\"ottingen, 37077 G\"ottingen, Germany}

\author{Peter Sollich}%
\affiliation{Institute for Theoretical Physics, University of G\"ottingen, 37077 G\"ottingen, Germany}
\affiliation{Department of Mathematics, King's College London, London WC2R 2LS, United Kingdom}

\author{Rituparno Mandal}
 \email{rituparno@rri.res.in}
\affiliation{Soft Condensed Matter Group, Raman Research Institute, Bengaluru 560080, Karnataka, India}

\begin{abstract}
All the fundamental interactions (such as gravity or electromagnetic interactions) are reciprocal in nature. However, in the macroscopic world, in particular outside equilibrium, non-reciprocal or non-mutual interactions are quite ubiquitous. Understanding the impact of such non-reciprocal interactions has drawn a significant amount of interest in physics and other fields of sciences in recent years. We explore a non-reciprocal version of coupled oscillators (known as the Kuramoto model) with the aim of understanding the role of non-reciprocity, particularly in relation to chimera states, where oscillators spontaneously break into mutually synchronous and asynchronous groups. 
Our findings suggest that non-reciprocity not only alters the state diagram of the chimera state significantly but can also lead to new dynamical states, such as traveling chimera, run-and-chase and coexistence phases.

\end{abstract}

\maketitle

\section{Introduction}

Synchronization is a widely observed phenomenon in both natural and synthetic systems~\cite{winfree1967biological, doi:10.1137/0150098, STROGATZ20001, rosenblum2003synchronization, kuramoto1984chemical}. Examples include synchronization of fireflies~\cite{buck1966biology}, collective motion of robots~\cite{giomi2013swarming}, 
Josephson arrays~\cite{WATANABE1994197},
pacemaker cells of the heart~\cite{pacemaker1984}, chirping of crickets~\cite{walker1969acoustic}, networks of neurons~\cite{doi:10.1137/0150098}, etc. In the context of dynamical systems, synchronization has been widely studied, and there exist a set of models such as the Kuramoto model~\cite{RevModPhys.77.137,kuramoto1984chemical}, the Sakaguchi-Kuramoto Model~\cite{sakaguchi1986soluble}, the Winfree model~\cite{winfree1967biological}, the Stuart Landau oscillators~\cite{CGLE}, pulse coupled oscillators (in neuroscience)~\cite{doi:10.1137/0150098}, which all display synchronization. Among such models of synchronization, the Kuramoto oscillator system~\cite{STROGATZ20001, rosenblum2003synchronization, PhysRevE.57.1563,RevModPhys.77.137,sakaguchi1986soluble, CGLE,kuramoto1984chemical,Rodrigues2016} 

has become the most extensively studied model. It captures the transition from an incoherent (unsynchronized) state to a synchronized state once the coupling strength exceeds a critical threshold, even when the oscillators start from random initial phases. These coupled oscillators can also spontaneously split into coexisting groups
of synchronized and unsynchronized oscillators and such states are known as chimera states~\cite{kuramoto2002coexistence, Abrams2004}. Over the years, the Kuramoto model has been extended in numerous directions and extensively studied~\cite{doi:10.1137/0150098,ARENAS200893,RevModPhys.77.137}, making it a cornerstone for understanding synchronization phenomena across physics, biology, and network science.

In this work, we report the effects of non-reciprocal couplings in Kuramoto oscillators, focusing on two canonical Kuramoto systems: $(a)$ two populations of oscillators with inter- and intra-population couplings, and $(b)$ oscillators arranged on a ring with distance dependent couplings.  
The reciprocal or symmetric versions of both systems have been extensively investigated in the past \cite{Abrams2004,Abrams2008}, especially in the context of chimera states. 

Recently, there has been a surge of interest in understanding the emergence and effects of non-reciprocal interactions in diverse range of non-equilibrium systems, where the action of one species/constituent on another is not balanced by an equal and opposite reaction~\cite{Ivlev2015,Fruchart2021Nonreciprocal}. Newton's third law is an example of a reciprocal interaction, which applies to the fundamental forces such as gravity and electromagnetic forces. Such action-reaction symmetry can also be observed in the case of effective forces in equilibrium ({\it{e.g.}}\ Casimir forces or depletion forces). Therefore, the dynamics of such systems is governed by a global free energy. Non-reciprocity, on the other hand, naturally arises in driven, active~\cite{JP2022unjamming, Ramaswamy2022} or living systems and leads to fundamentally new collective phenomena~\cite{Ivlev2015,   Markovich2021, Yuan2023,Fruchart2023OddViscosityElasticity,Markovich2024Nonreciprocity,gu2025emergence}.

Non-reciprocal couplings have been shown to generate novel dynamical states in many-body systems, breaking time-reversal symmetry and enabling directional transport. Experiments and theoretical studies across chemical droplets, active particles, social dynamics, oscillator networks, and spin systems highlight non-reciprocity as a key ingredient for emergent phases, instabilities and sustained oscillations \cite{Meredith2020,You2020NonreciprocityTraveling, Loos2022, mandal2024learning, Mandal24, Sociohydrodynamics2025, Avni25, NRSG2025}. Similar concepts ({\textit{i.e.}}\ breaking of action-reaction symmetry), once extended to colloidal assemblies, nanoparticle systems and continuum descriptions of active matter, lead to rich pattern formation and transport phenomena, and exotic time dependent states~\cite{Saha2020,Osat2022NatNano, Cocconi2023PRResearch, Duan2023PRL,Dinelli2023NatComm, Brauns2024PRX,Osat2024PRL,  AlHarraq2025AdvSci}.

Although prior studies ~\cite{Hong2011, Ivlev2015, Fruchart2021Nonreciprocal, Asym_Kuramoto_PRE_2024, sander2024kuramoto} have examined some aspects of how asymmetry or non-reciprocity in interactions can affect collective dynamics in coupled oscillators there is, to our knowledge, no systematic study of the effect of non-reciprocity in coupled oscillators, specifically in those models that were designed to study chimera states. Combining extensive numerical simulations and mean-field-like analytical approaches, we show that allowing non-reciprocity in pairwise couplings can fundamentally reshape stability regimes, give rise to new sequences of transitions, and produce novel time-dependent states when compared to the previously studied symmetric models~\cite{Abrams2008,Abrams2004}. The two-population model of Ref.~\cite{Abrams2008} and the  spatial extended model described in Ref.~\cite{Abrams2004}, provide a minimal setting for isolating how non-reciprocity impacts chimera states in Kuramoto oscillator model. We also observe the emergence of time-dependent dynamical states in the non-reciprocal version of the spatially extended model, like the run-and-chase phase, traveling chimera, and spatial coexistence of run-and-chase and synchronized phases, which we will refer to as a chimera state of type II.

The paper is organized as follows. In Section II, we introduce the two-population model with nonreciprocal inter-population coupling and report the main results from our numerical simulations. We show that non-reciprocity reshapes the state diagram, in particular the regions of stable and breathing chimera states, and also gives rise to a ``run-and-chase'' phase, which we analyze and summarize in a phase diagram
in the coupling strength ($A$) and non-reciprocity ($\delta$) plane. We derive the corresponding mean field equations, draw the bifurcation curves and match them with the phase diagram obtained from the simulations. In Section III we extend the analysis to a spatially distributed version of the model and numerically construct the $(A,\delta)$-phase diagram, identifying five distinct phases: synchronized, chimera, traveling chimera, run-and-chase, and a chimera state of type II. We summarize our findings and provide an outlook in Section IV.

\section{Two population model}
\subsection{Model and Numerical Details}

\begin{figure}
    \centering
\includegraphics[width=0.9\linewidth]{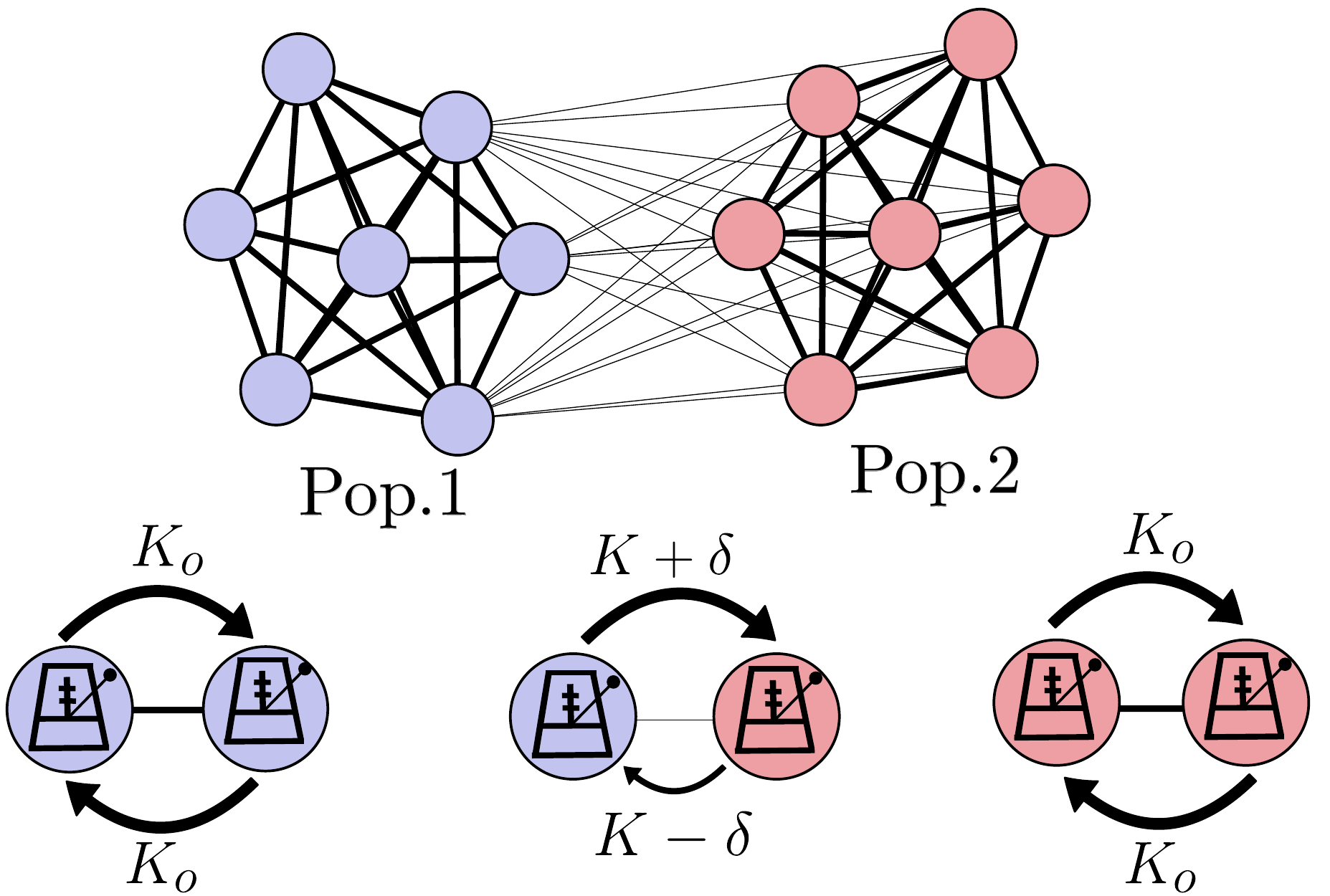}
    \caption{A schematic of our model showing two populations of oscillators with all-to-all coupling. The intra-population coupling has strength $K_0$ whereas non-reciprocal interaction have been implemented in the inter-population coupling as $K^{\rm NR}_{12}=K + \delta$ and $K^{\rm NR}_{21}=K - \delta$. 
    }
    \label{Fig1}
\end{figure}

Following Ref.~\cite{Abrams2008}, we consider two groups or populations of identical oscillators (see Fig.~\ref{Fig1} for a schematic), with intra- and inter-population couplings. The equation for the evolution of the oscillator phases is then given by 
\begin{equation}
\frac{d\theta^{\sigma}_{i}}{dt} = \omega + \sum_{\sigma^{\prime}=1}^{2} \frac{K^{\rm NR}_{\sigma\sigma^{\prime}}}{N_{\sigma^{\prime}}} \sum_{j = 1}^{N}\sin(\theta_{j}^{\sigma^{\prime}} -  \theta_{i}^{\sigma} - \alpha),
    \label{eq1}
\end{equation}
where $\theta_{i}^{\sigma}$ is the phase of the $i$-th oscillator belonging to population $\sigma$ (with $\sigma \in \{1,2\}$), $\omega$ is the natural frequency of the oscillators and $\alpha$ is the phase lag between the oscillators of the two populations, also known as the disorder parameter. This model describes an all-to-all coupled set of oscillators.

In Eq.~\ref{eq1} $K^{\rm NR}_{\sigma \sigma'}$ (with $\sigma,\sigma' \in \{1,2\}$) is the coupling or interaction matrix between the population $\sigma$ and $\sigma'$. We modify the interaction matrix given in Ref.~\cite{Abrams2008} by introducing a non-reciprocity parameter $\delta$,
\begin{equation}
    K^{\rm NR} \equiv K^{\rm R} + \begin{pmatrix}
        0 &   \delta\\  -\delta & 0 \end{pmatrix} \equiv \begin{pmatrix}
       K_{11} &   K_{12}+\delta\\ K_{21} -\delta & K_{22}
         \end{pmatrix}
    \label{eq2}
\end{equation}
where $K_{11}$ and $K_{22}$ represent intra-population couplings and $K_{12}=K_{21}$
represents the inter-population coupling of the reciprocal model~\cite{Abrams2008}. Following Ref.~\cite{Abrams2008}
we further assume the populations are identical  $K_{11}=K_{22}=K_0=\frac{1+A}{2}$ and write their influence on each other as $K_{12}=K_{21}=K=\frac{1-A}{2}$ in the reciprocal case, where
$A$ is a control parameter that determines the relative strength of the intra-population and inter-population coupling.

For the numerical simulations, the parameter $\omega$ is taken to be the same for both the populations (since they are identical) and has been set to zero without any loss of generality (see Ref.~\cite{RevModPhys.77.137} or Supplementary Information for details). 
We set $\alpha=1.47$ for most of our study; for a set of other values of $\alpha$ we report the results in the Supplementary Information. We explore values of $A$ in the range of 0 to 0.5, while we parameterize $\delta$ in terms of $\Delta = \sgn(\delta) \ln(1 + \frac{|\delta|}{\epsilon})$. We set $\epsilon =  0.003$ and vary $\Delta$ between $-6$ and $6$. Note that we mostly refer to $\delta$ itself as the non-reciprocity parameter, but use $\Delta$ to cover a logarithmic scale across both positive and negative values of $\delta$, and to draw the corresponding phase diagrams.

We numerically solve the system of coupled ODEs, Eq.~\ref{eq1}, using the fourth order Runge-Kutta method. The simulations are done for a system of  $N_1=N_2=128$ oscillators using an adaptive step size algorithm until time $t = 5000$. As the basin of attraction of the chimera state is smaller than that of the synchronized state~\cite{SarahLoos2016}, following ~\cite{Abrams2004, Abrams2008,Abrams2013,Abrams2015}, we start from a chimera-like initial condition and observe the resulting behavior of the dynamical system. We report our main observations in the following section.
 
\subsection{Results}
\subsubsection{Alterations in the Transition Points}

\begin{figure}[]
\centering
\includegraphics[width = .88\columnwidth]{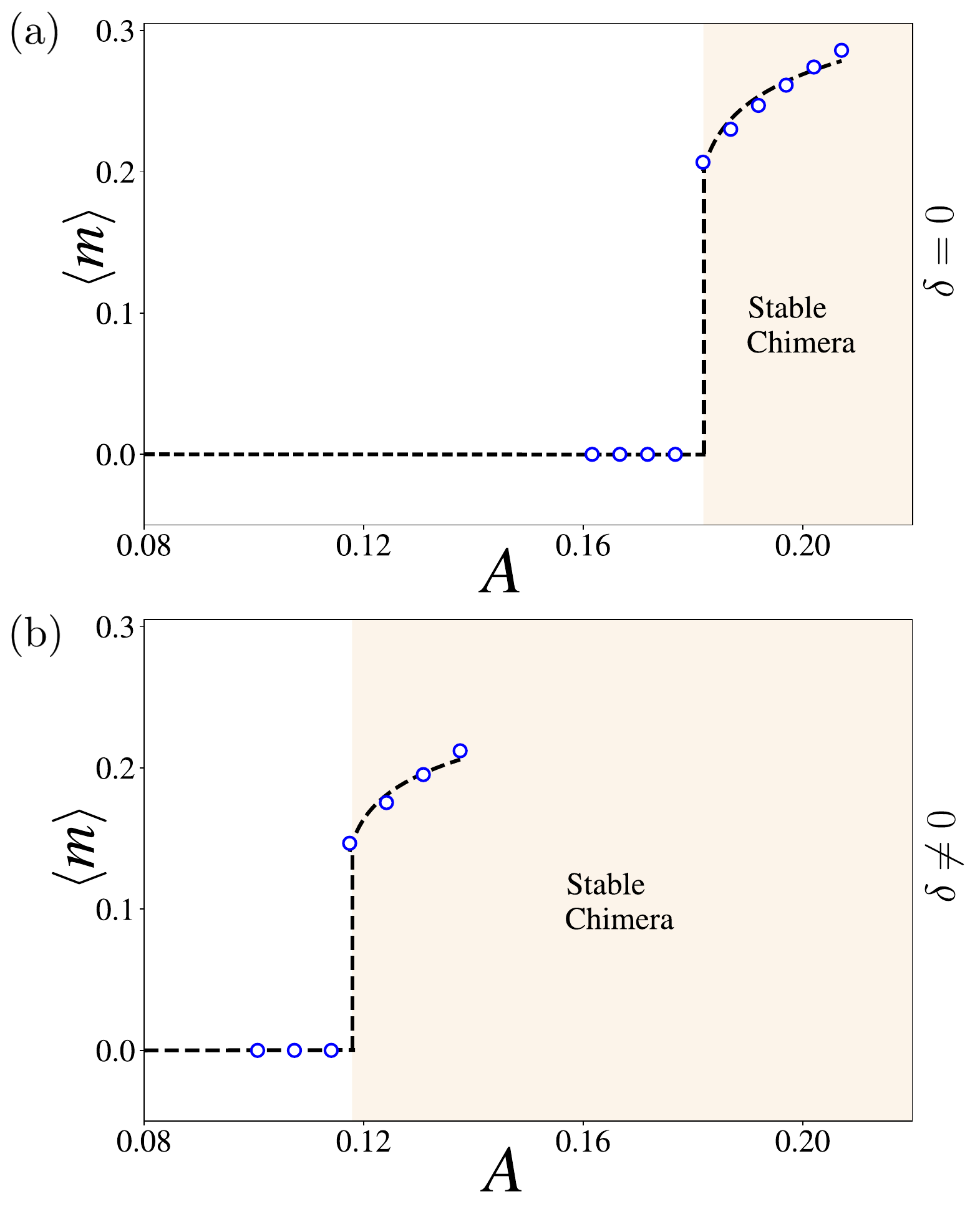}
\caption{(a) Order parameter $\langle m \rangle$ showing a transition from the synchronized state to the stable chimera for $\delta = 0$, at a critical value $A_{c1} = 0.18$. (b) Similar plot for $\delta=0.005$, where the transition occurs at $A_{c1} = 0.118$. The blue circles represent simulation data and the dashed line is a fit to the functional form $\langle m \rangle - m_0 \propto |A-A_{c1}|^{\gamma}$, where $m_{0}$ is the order parameter just above the transition and $\gamma$ is the scaling exponent.}
\label{Fig2}
\end{figure}

We start with the question whether the states ({\it{e.g.}}\ synchronized, disordered, chimera) and the corresponding transition points observed in the reciprocal model remain intact after the incorporation of the non-reciprocity parameter ($\delta$) or not.
The synchronization is often characterized by an order parameter $r_{\sigma}(t)$~\cite{STROGATZ20001,Abrams2008,Abrams2004,kuramoto2002coexistence}, where $\sigma$ is the population index and $\sigma \in 1,2$. The synchronization order parameter $r_{\sigma}(t)$ measures the instantaneous average of the phases of the oscillators in population $\sigma$. Explicitly we write for population $\sigma$, containing $N_\sigma$ oscillator,
\begin{equation}
    r_{\sigma}(t) = \frac{1}{N_{\sigma}} \Biggl|\sum_{j=1}^{N_{\sigma}}e^{i\theta_{j}^{\sigma}(t)}\Biggr|\ .
    \label{eq3}
\end{equation}
When the oscillators are completely synchronized one obtains $r_{\sigma}(t) = 1$ while random desynchronized phases give $r_\sigma(t) = 0$; values $0<r_\sigma(t)<1$ then indicate partial synchronization. The initial conditions of our simulations are always chimera states: we start from the state where  population 1 is in complete synchronization 
while the population 2 is relatively disordered. We observe that the initially synchronized population remains synchronized 
for the simulation time period. Therefore the relevant synchronization order parameter is the one that describes the initially disordered group; we abbreviate this by $r(t)$. We also define $m(t) = 1-r(t)$ and $\langle m \rangle=\frac{1}{T}\int^T_0 m(t) dt$ where $T$ is the time period over which the averages are computed in the steady state. Therefore, $\langle m \rangle$ is zero in the completely synchronized state and a non-zero $\langle m \rangle$ indicates a chimera state (for our initial condition $r(0)=1$ for population 1 and $r(0) \sim 0.7$ for population 2).

\begin{figure}[]
\centering
\includegraphics[width = .86\columnwidth]{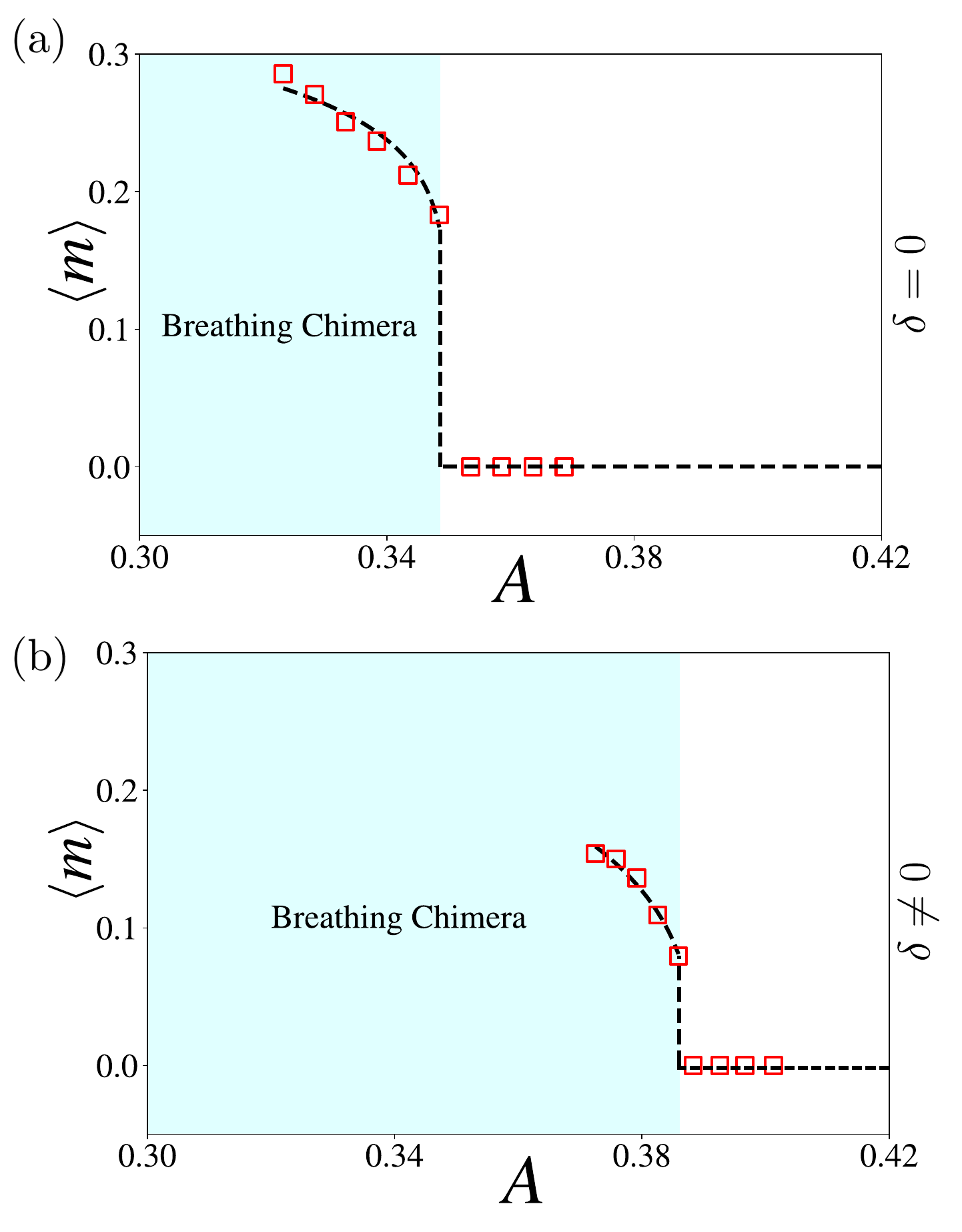}
\caption{(a) Order parameter $\langle m \rangle$ showing a transition at $A_{c2}=0.353$  from the breathing chimera to the synchronized state for $\delta = 0$. (b) A similar plot for  $\delta=0.005$ shows a transition at $A_{c2} = 0.39$. The red squares are data points obtained from simulations and the dashed line is a fit to the functional form $m - m_0 \propto |A-A_{c2}|^{\gamma}$.
}
\label{Fig3}
\end{figure}

In Fig.~\ref{Fig2}(a) we show the variation of the order parameter $\langle m \rangle$ as a function of $A$, showing a transition from a synchronized state to a stable chimera state (where the order parameter remains constant at some  $r(t)<1$ and does not fluctuate/oscillate with time) for the reciprocal case $\delta=0$. This transition can be captured by the form 
\begin{equation}
    \langle m \rangle  - m_{0} \propto \lvert A - A_{c1}\rvert^{\gamma}
    \label{eqtrans}
\end{equation}
(dashed line in Fig.~\ref{Fig2}) where $A_{c}$  represents the transition point, $\gamma$ is a scaling exponent and $m_{0}$ is the value of $\langle m \rangle$ directly above $A_{c1}$. For the non-reciprocal case, with $\delta= 0.005$, we see a qualitatively similar transition -- {\em i.e.}\ power law behavior beyond a first order transition -- between the synchronized state and a stable chimera, but happening at a lower value of $A_{c1}$ (see Fig.~\ref{Fig2}(b)). 

Next, we examine the transition from the breathing chimera (where the order parameter $r(t) < 1$ but its value oscillates or `breathes' with time) to the synchronized state. Fig.~\ref{Fig3}(a) shows such a transition as a function of $A$ for the reciprocal case $\delta=0$, which can 
again be fitted by the form of Eq.~\ref{eqtrans}; we denote the value of $A$  for this transition by $A_{c2}$.
In the presence of non-reciprocity, $\delta = 0.005$, we observe a similar transition
but at a higher $A_{c2}$, see Fig.~\ref{Fig3}(b).

From Fig.~\ref{Fig2} and Fig.~\ref{Fig3}, it is quite evident that the introduction of non-reciprocity leads to significant alterations in the transition points $A_{c1}$ and $A_{c2}$ for both types of transition, {\it{i.e.}}\ going with increasing $A$ from the synchronized state to the stable chimera state, and from the breathing chimera state to the synchronized state. The transition from the synchronized state to the stable chimera state occurs at a lower value of $A_{c1}$, and that of the breathing chimera to the synchronized state happens at a higher value of $A_{c2}$ when a small amount of non-reciprocity is introduced. Therefore, the transition points can be easily tuned by changing the strength of non-reciprocity $\delta$. We explore this in more detail in the \textit{Phase Diagram} section below. 
\begin{figure}[]
\centering
\includegraphics[width = 1\columnwidth]{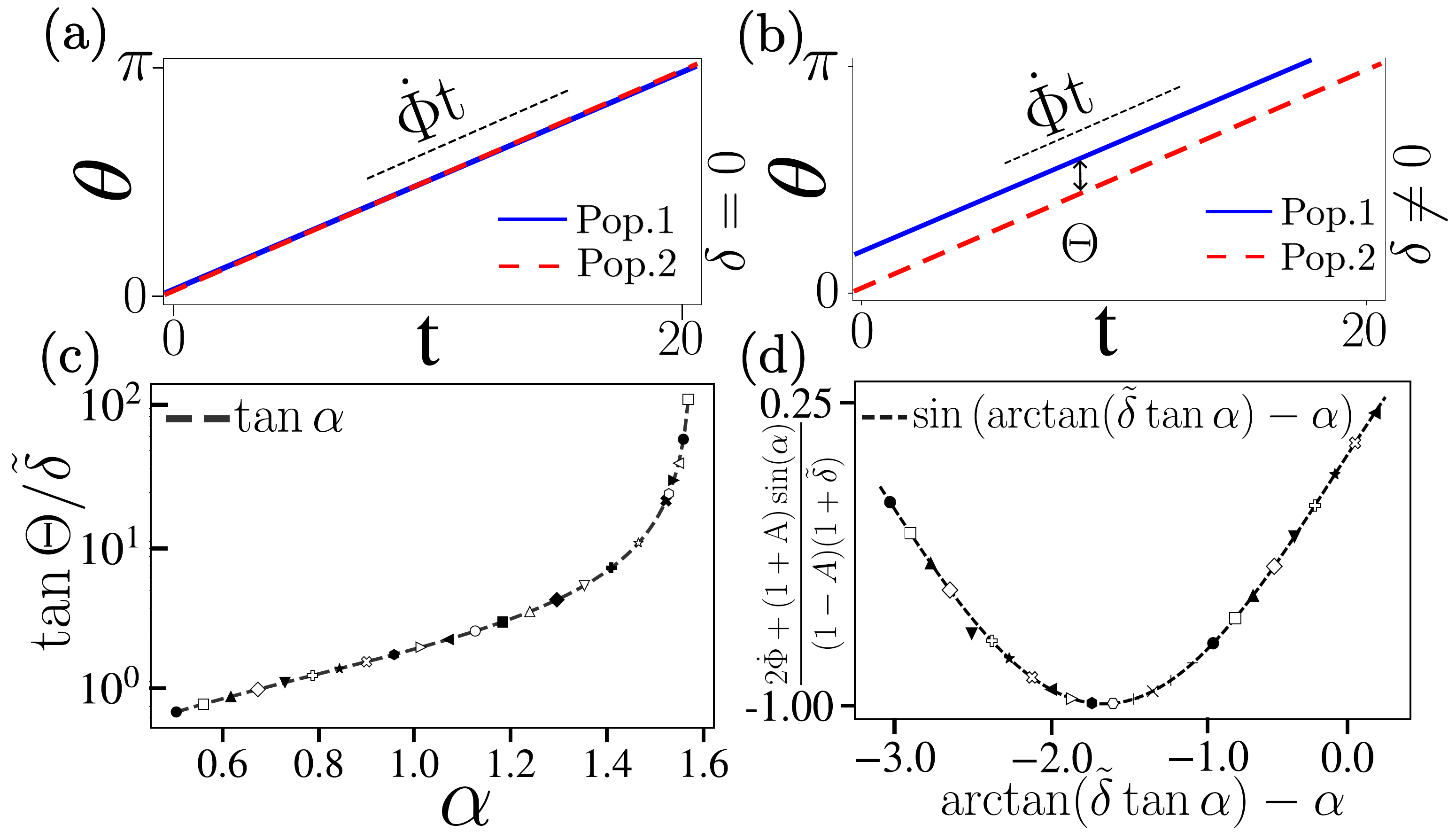}
\caption{(a) Mean phase $\theta_\sigma$ ($\sigma \in 1,2$) for the reciprocal case, $\delta = 0$ shows the phase difference ($\Theta$) between the two populations to be zero. (b) For the non-reciprocal case, $\delta=0.6$, though the oscillators in each population have individually synchronized with each other, there exists a finite phase difference $\Theta$, between the oscillators of the two populations. (c,d) show the comparison between the analytical prediction (black dashed line) and the simulation results (symbols) for different $\delta, \alpha$ and $A$.}
\label{Fig4}
\end{figure}

\subsubsection{Run-and-Chase}

Apart from modifying the region of the chimera phase in the parameter space, non-reciprocity also introduces a time-dependent state, which we call the run-and-chase state following Ref.~\cite{Fruchart2021Nonreciprocal}. A generic feature of non-reciprocal systems is that instead of having a global free energy function, the non-reciprocity creates conflicting objectives within its sub-systems~ \cite{Sompolinsky1986,%
Fruchart2021Nonreciprocal,%
Zhang2023NonReciprocalEntropy,
mandal2024learning,Mandal24,Avni25}, in a way that is analogous to a predator chasing its prey while the prey attempts to evade it. In the context of coupled oscillator populations, introducing non-reciprocity creates an asymmetric interaction: for $\delta>0$ 
oscillators in population $1$ are more strongly coupled to population $2$, attempt to synchronize with them, while population $2$, which is more weakly influenced in return, tends to drift out of sync. At the same time, oscillators within each population remain synchronized with one another. 

Fig.~\ref{Fig4}(a) shows the time evolution of the mean phases of the oscillators of both the populations for the reciprocal case, and Fig.~\ref{Fig4}(b) for non-reciprocal inter-population interactions. Here, $\theta_{\sigma} = \sum_{j=1}^{N} \theta_{j}^{\sigma}$, where $\sigma$ is the population index. Note that in Fig.~\ref{Fig4}(a,b) we have dropped the population index for visual ease and instead use two different colors and line styles to indicate the two populations. From Fig.~\ref{Fig4}(a,b) we observe that non-reciprocity introduces a stable and finite phase difference $\Theta=\theta_1-\theta_2$ between the populations. Further inspection also shows that a nonzero $\delta$ changes the speed of the mean angle $\frac{d\Phi}{dt}$ where $\Phi=\frac{\theta_1+\theta_2}{2}$. 
We will now try to understand the dependence of $\Theta$ and $\omega$ on the system parameters such as $A, \alpha$ and $\delta$. Note that we have varied $\alpha$ in this section, though for most of the results presented in the main text we don't vary $\alpha$.

To analyze this run-and-chase state, we start by assuming that the oscillators in both populations are fully synchronized, $\theta_{j}^{\sigma}= \theta_{\sigma}$ for $\sigma \in \{1,2\}$ for all $j$. We then substitute the phases of two fully synchronized populations ($\theta_1$ and $\theta_2$) into Eq.(1) and write down the equations for the time evolution of $\Theta$ and $\Phi$, the two global coordinates that encode respectively the phase difference ($\theta_2-\theta_1$) and the mean angle ($\frac{\theta_1+\theta_2}{2}$). The dynamics of $\Theta$ and $\Phi$ can be written as 
\begin{equation}
    \frac{d\Theta}{dt} = -2\delta \sin \alpha \cos \Theta + (1-A)\cos\alpha \sin\Theta
    \label{dthetat}
\end{equation}
and 
\begin{equation}
    \dot{\Phi} = \frac{1+A}{2} \sin{(-\alpha)} + (\frac{1-A}{2} + \delta)\sin(\Theta-\alpha)
    \label{Phidot}
\end{equation}
For $\delta = 0$, the only stable fixed point is $\Theta = 0$, corresponding to perfect synchrony. However, any finite non-reciprocity ($\delta \neq 0$) shifts the fixed point to a nonzero $\Theta$ that reflects persistent pursuit and escape between the populations. Explicitly one finds
\begin{equation}
\Theta = \arctan \left(\tilde{\delta}\tan\alpha \right)
 \label{eq5}
\end{equation}
where $\tilde{\delta}=2 \delta/(1-A)$. Inserting this value of $\Theta$ into 
Eq.~\ref{Phidot} we obtain an equation determining $\dot{\Phi}$:
\begin{equation}
   \sin \left( \arctan \left({\tilde{\delta}} \tan\alpha \right)-\alpha \right) =\frac{2 \dot{\Phi} + (1+A) \sin\alpha}{(1-A)(1+\tilde{\delta})}.  
    \label{eq6}
\end{equation}
Figs.~\ref{Fig4}(c),(d) show a collapse of $\Theta$ and $\frac{d\Phi}{dt}$, respectively, for different values of $\alpha$, $\delta$, and $A$ (for the details of the calculation see Supplementary Information) and demonstrate the usefulness of the effective two oscillator model from above. 

\subsubsection{Phase Diagram}

\begin{figure}[]
\centering
\includegraphics[width = \columnwidth]{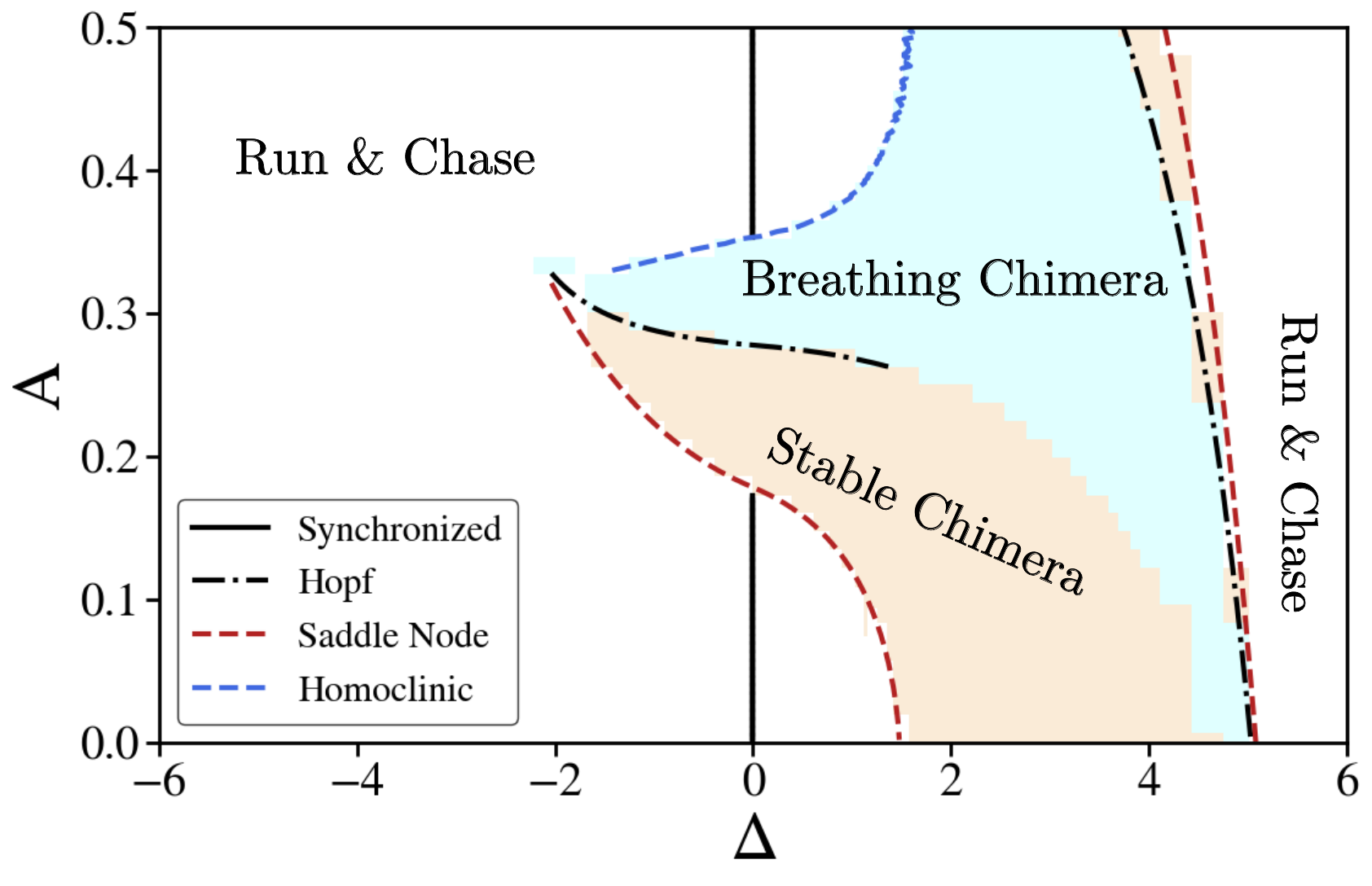}
\caption{Phase diagram of the stability of states in the $(A,\delta)$ plane, constructed by running simulations with $\alpha=1.47$. The value of $\delta$ is represented on the $x$-axis in terms of $\Delta = \sgn(\delta) \ln(1 + \frac{|\delta|}{\epsilon})$ with scale parameter $\epsilon= 0.003$. A saddle-node bifurcation occurs at the transition from the run-and-chase phase to the stable chimera phase (red dashed lines) and the transition from stable chimera to breathing chimera is a Hopf bifurcation (black dashed-dot lines). A homoclinic bifurcation occurs when the breathing chimera transitions into the run-and-chase phase (marked by the blue dashed line). All the bifurcation lines are obtained theoretically;  see text for details.} 
\label{Fig5}
\end{figure}

To get a global understanding of the system behavior we carried out numerical simulations for the dynamics of the oscillators, for fixed $\alpha=1.47$ as before but with both $A$ and the non-reciprocity parameter $\delta$ varying across a broad range. We analyze the order parameter $r(t)$  (for population 2; population 1 always ordered)
in their steady state and classified the system into one of the four dynamical states, namely synchronization, stable chimera, breathing chimera, or run-and-chase. If the time average of the order parameter $\langle|r(t)|\rangle_{t}=1$ 
at the steady state, then the populations are synchronized. If $\langle|r(t)|\rangle_{t}<1$ and variance is $0$, the system is defined to be in a stable chimera state, while for variance $> 0$, the system is defined to be in a breathing chimera state. 

Fig.~\ref{Fig5} shows the different phases in the $A$ and $\Delta$ plane, where $\Delta = \sgn(\delta) \ln(1 + \frac{|\delta|}{\epsilon})$, where $\epsilon$ is a scaling parameter (here, $\epsilon =  0.003$).
It is quite evident from the phase diagram that non-reciprocity influences the region where chimera states exist. In the Supplementary Information we provide figures showing the phase diagram for a few other values of $\alpha$, which show qualitatively similar behavior.

In Fig.~\ref{Fig5}, the solid
line at $\Delta = 0$ identifies the synchronized phase (for $A<A_{c1}$ and $A>A_{c2}$). 
As discussed in the previous section, when non-reciprocity is introduced ($\delta \neq 0$) the fully synchronized state gives way to a run-and-chase phase. 
As mentioned before, in this regime, the two populations behave like a predator and a prey: population 1 continuously tries to synchronize with population 2, while population 2 drifts away. Despite this run-and-chase dynamics between the populations, oscillators within each population remain mutually synchronized. 

As demonstrated earlier, in Fig.\ref{Fig2} and \ref{Fig3} the critical values of the coupling parameters $A_{c1}$ and $A_{c2}$, at which the system transitions into the stable chimera and transitions out of the breathing chimera phases are modified under the introduction of non-reciprocity. The phase diagram (see Fig.\ref{Fig5}) further indicates that the transition point from the stable to the breathing chimera is also dependent on the strength of non-reciprocity $\delta$ (or equivalently $\Delta$). Moreover, non-reciprocity gives rise to an additional qualitative feature not present in the reciprocal ($\textit{i.e.,}$ the symmetric coupling) case, the reentrant transition with respect to $A$. For instance, at a fixed value of $\Delta \simeq 4.2$, varying $\text{A}$ leads to the sequence of transitions from the stable chimera to breathing chimera and again to the stable chimera, demonstrating a reentrant recovery of the stable chimera phase at a higher $\text{A}$ value. The bifurcation lines (between different chimera states and other states) in the phase diagram are obtained analytically from the mean field equations discussed in the following section. 

\subsubsection{Mean Field Analysis}

To obtain the phase boundaries we use mean field equations, considering the limit of $N_\sigma\to\infty$, which leads to the continuity equation~\cite{Abrams2008}
\begin{equation}
    \frac{\partial f^\sigma}{\partial t} + \frac{\partial}{\partial\theta} (f^\sigma v^\sigma) = 0
    \label{continuity}
\end{equation}
where $f^\sigma(\theta,t)$ is the probability density of population $\sigma$ so that $f^\sigma(\theta,t)d\theta$ is the probability for an oscillator in that population to have a phase angle between $\theta$ and $\theta+d\theta$. The factor $v^\sigma(\theta,t)$ is the drift term, which can be obtained from the continuum limit of Eq.~\ref{eq1} 
for the phase of a single oscillator as 
\begin{equation}
    v^\sigma(\theta,t) = \omega +\sum_{\sigma\sigma'} K^{\rm NR}_{\sigma\sigma'} \int \sin (\theta'-\theta-\alpha) f^{\sigma'}(\theta',t) d\theta'
\end{equation}
From here on we follow the methodology used in Ref.~\cite{Abrams2008}. We only outline the steps here and refer to the Supplementary Information for details. Into the continuity equation, Eq.~\ref{continuity} 
, we introduce the Ott-Antonsen {\it{ansatz}}~\cite{Ott2008} for the probability density function $f^{\sigma}$, which reduces Eq.~\ref{continuity} to a single amplitude equation. We then write the complex amplitude 
in polar coordinates as $\rho_\sigma e^{-i\phi_\sigma}$ to get equations for two real variables corresponding to the modulus $\rho_\sigma$ and the phase $\phi_\sigma$ of the complex amplitude. In the analysis of chimera states we have one population (population 1) of fully synchronized population of oscillators and therefore set the modulus of the corresponding complex amplitude of unity {\it{i.e.}} $\rho_1=1,\rho_2=r$.
Finally, we can write the equations characterizing the chimera state as\begin{eqnarray}\label{2variable_reduced_eqn}
    \dot{r} &=& \frac{1-r^2}{2}\left(\frac{1+A}{2} r \cos\alpha+\left(\frac{1-A}{2}-\delta\right)\cos(\Theta-\alpha)\right)\cr
    \dot{\Theta} &=& \frac{1+r^2}{2r}\left[ \frac{1+A}{2} r\sin\alpha -\left(\frac{1-A}{2}-\delta\right)\sin(\Theta-\alpha) \right]\cr
    &-&\frac{1+A}{2} \sin\alpha -\left(\frac{1-A}{2}+\delta\right) r\sin(\Theta+\alpha)
\end{eqnarray}
where 
$\Theta$ is the difference between the phases of the complex amplitudes for the two populations. 

To compare the mean field analysis with the simulation results, we fix $\alpha=1.47$. 
For stable chimera states we solve the condition of stationary $r$, $\dot{r}=0$, we can express the corresponding $A^*$ in terms of $(r^*,\delta^*, \Theta^*)$. 
Using the result in the condition  $\dot{\Theta}=0$ we then obtain an expression of $r^*$ in terms of $(\delta^*, \Theta^*)$, and from there also $A^*$ in terms of the same two variables.
To locate bifurcations of Hopf and saddle-node type, we analyze the behavior around the stable chimera fixed points parametrized in terms of $(\delta^*, \Theta^*)$. This is done by linearizing the equations near these fixed points and analyzing the properties of the corresponding Jacobian matrix $J$ with $J_{11}=\frac{\partial\dot{r}}{\partial r}$, $J_{12}= \frac{\partial\dot{r}}{\partial \Theta}$, $J_{21}=\frac{\partial\dot{\Theta}}{\partial r}$ and $J_{22}=\frac{\partial\dot{\Theta}}{\partial \Theta}$.  

A Hopf bifurcation occurs when a fixed point gives rise to a limit cycle while changing parameters. This is indicated by the eigenvalues of $J$ becoming purely imaginary, {\it{i.e.}} $\text{Tr}(J)=0$. By solving for $\text{Tr}(J)=0$ using the fixed point solutions $(A^*,r^*)$, we obtain the Hopf bifurcation line in the $(\delta, \Theta)$ 
plane, from which the corresponding line in the ($\delta,A$) can be found.
A saddle-node bifurcation happens at a point when one of the eigenvalues of $J$ vanishes, {\it{i.e.}} $\text{Det}(J)=0$, while changing the system parameters. We obtain these bifurcation lines following the same procedure as for the Hopf case.
Homoclinic bifurcations, finally, occur when a limit cycle solution and a saddle-node solution meet and annihilate each other. As this is a global bifurcation, it cannot be obtained by linearization of the dynamical equations for $r$ and $\Theta$. To obtain the homoclinic bifurcation line, we therefore solve these equations numerically. The results for all bifurcations are summarized in Fig.~\ref{Fig5} and show very good agreement with our direct numerical simulations. 

\section{ Spatially extended coupled oscillator model}

\subsection{Model and Numerical Details}

\begin{figure}
\centering\includegraphics[width=0.9\linewidth]{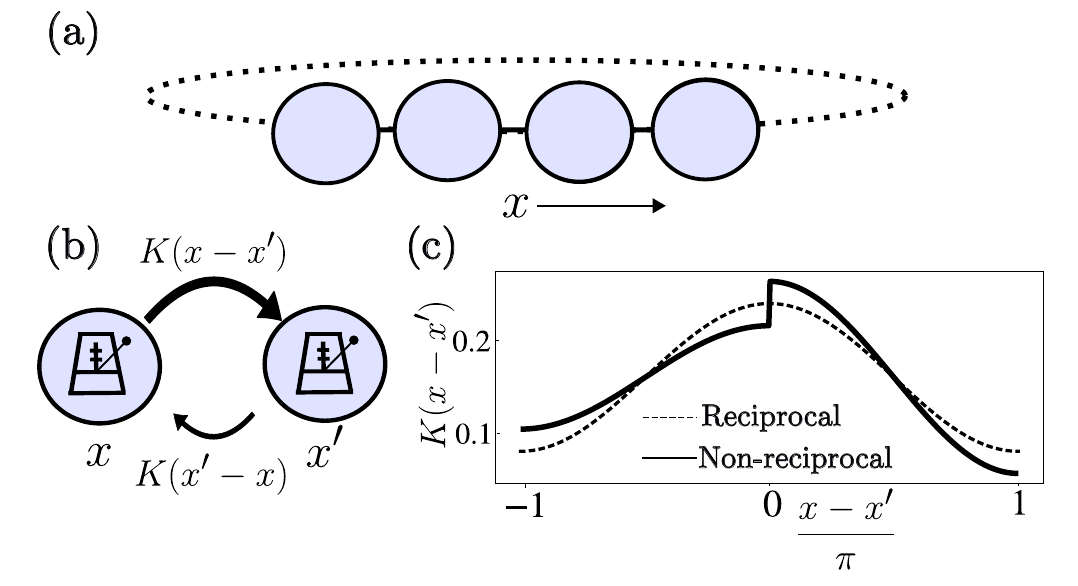}
    \caption{Oscillators embedded in space with non-local, non-reciprocal interactions (in the schematic, an oscillator is more strongly coupled to the right neighbor than the left one). }
    \label{Fig6}
\end{figure}

In this section we move to another coupled oscillator model where the oscillators are arranged in one dimension and have interactions dependent on their separation; for the simpler case of symmetric, {\em i.e.}\ reciprocal, interactions this model was studied in Ref.~\cite{Abrams2004,kuramoto2002coexistence}). 
Moving directly to a continuum description, the time evolution of the phases of the oscillators is given by 
\begin{equation}
    \frac{\partial \theta}{\partial t} = \omega + \int_{-\pi}^{\pi} K(x-x^{\prime}) \sin[\theta(x^{\prime},t) - \theta(x,t) + \alpha] dx^{\prime},
    \label{eq14}
\end{equation}
where $\theta(x,t)$ denotes the phase of the oscillator at position $x\in(-\pi,\pi)$ 
and at time $t$, $\omega$ is the natural frequency as before and $\alpha$ is the phase lag parameter. To avoid boundary effects we consider periodic boundary conditions. This means that the oscillators are arranged in a ring, as shown in Fig.~\ref{Fig6}. The form of the interaction for the symmetric case is given by \cite{Abrams2004} 
$K(x-x') = \frac{1}{2\pi}(1+A\cos{(x-x')})$ where $A$ represents the coupling strength. Note that the cosine function makes the interaction symmetric under the interchange $x-x'\leftrightarrow x'-x$, which ensures reciprocity. 

Some previous works on spatially extended versions of this model~\cite{Smirnov2024,TCWeakTemporal} have introduced asymmetry by skewing the coupling kernel, typically in the form $K(x-x') = \frac{1}{2\pi}(1+A\cos{(x-x')} + B\sin{(x-x')})$, where the third term tilts the interaction profile and breaks the right-left ($x - x^{\prime} \leftrightarrow x^{\prime} - x$) symmetry. Our construction is slightly different: we impose an explicit left–right asymmetry by making each oscillator more strongly coupled to its right neighbor than to its left,
\begin{equation}
K(x-x') = \frac{1}{2\pi}(1+(A+\delta \sgn(x-x'))\cos{(x-x')}).
\label{kernel}
\end{equation}
This choice directly breaks reciprocity $K(x-x^{\prime})\neq K(x^{\prime}-x)$ between two oscillators located at $x$ and $x'$, as shown in Fig.~\ref{Fig6}(b,c). In this construction $\delta$ controls the degree of non-reciprocity in the model. 
Note that the sign function is evaluated by taking for $x'$ the periodic image closest to $x$, so that $\sgn(x-x')$ changes sign both at $x=x'$ and at $x=x'\pm \pi$.

\subsection{Results}
\begin{figure*}[!ht]
    \centering
    \includegraphics[width = 2\columnwidth]{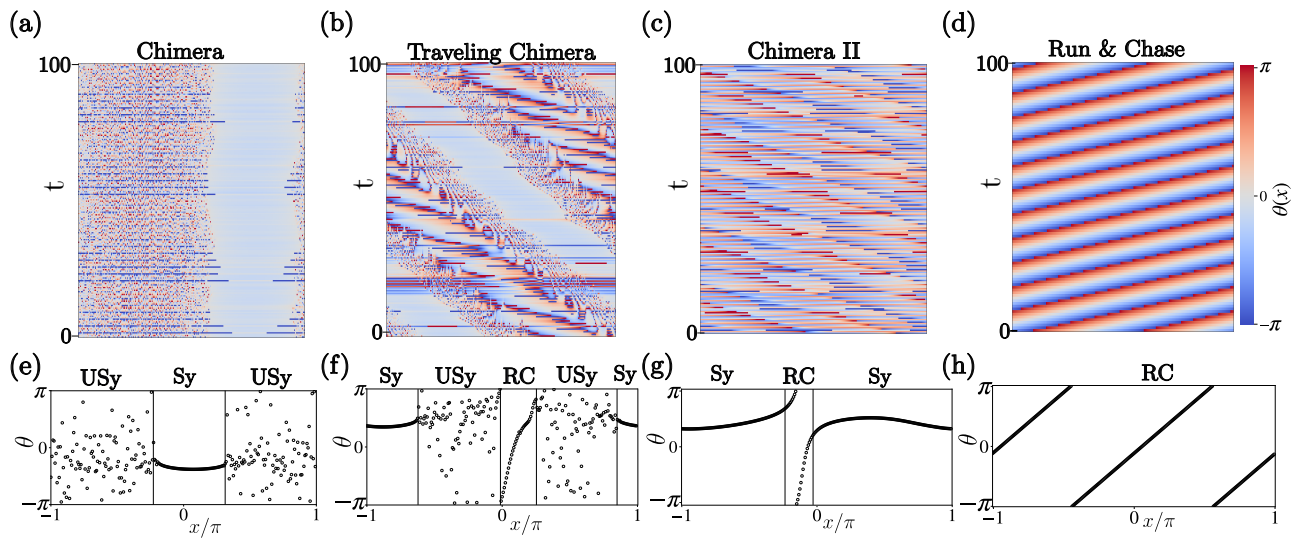}
    \caption{All the kymographs have the $x$-axis denoting space, running from $-\pi$ to $\pi$ with periodic boundary conditions. The $y$-axis indicates time and the color represents the $\theta$ values. (a) Kymograph displaying the chimera phase obtained for $A = 0.7$ and $\delta=0.001$. (b)  Kymograph for a traveling chimera phase, obtained for $A=0.7$ and $\delta = 0.057$. Here, the regions of synchronized and unsynchronized oscillators are no longer stationary in time. (c) The chimera of type II, obtained for $A=0.75$, $\delta=0.212$. It shows coexistence of the synchronized phase and the run-and-chase phase. (d) Kymograph of the run-and-chase phase, obtained at $A=0.5$, $\delta=0.3$. (e,f,g,h) show the instantaneous phase profile of the chimera phase, traveling chimera phase, chimera of type II, and the run-and-chase phase, respectively. The $x$-axis denotes space (scaled by $\pi$) and the y-axis gives the phase of the oscillators at $x$ at the chosen instant in time. The spatial profiles are measured at $t = 25000$, after each system has reached its steady state. 
    The different spatial regions occupied by synchronized (Sy), unsynchronized (USy) and run-and-chase (RC) phases are labeled for clarity.}
    \label{Fig7}
\end{figure*}
We numerically integrated Eq.~\ref{eq14} using a fourth-order Runge-Kutta method 
with a fixed time step of $dt=0.025$. The simulations were run for a total time $t \sim 4 \times 10^4$. The spatial domain $(-\pi, \pi)$ was discretized into $N_g=256$ grid points and the interaction kernel $K(z)$ was implemented via the convolution theorem using the Fast Fourier Transform (FFT), which significantly reduces computational cost from $O(N^{2})$ to $O(N\log(N))$~\cite{press2007numerical}. The initial values of the phases $\theta(x)$ were sampled from a zero-mean and unit-variance normal distribution.  The value of $\omega$ was chosen to be zero without loss of generality, as in the all-to-all model. In the next section, we discuss the results of the simulation.

\subsubsection{Synchronization}

When the coupling strength $A$ is below some critical value $A_c$ (for example, $A_c \approx 0.64$ for $\alpha = 1.43$ and $\delta=0$), then the oscillators tend to synchronize with their neighbors, and this eventually results in complete synchronization of the system. From our simulations it appears that the introduction of the non-reciprocity parameter $\delta$ does not affect synchronization unless $\delta$ exceeds a critical value, at which the system evolves into a coexistence 
between a synchronized and a run-and-chase phases; the latter is discussed in detail below. The synchronized phase is stable for $A<A_c$ for the reciprocal case and has been discussed in detail in Ref.~\cite{Abrams2004}. 

\subsubsection{Chimera}

The chimera state in this model is characterized by the coexistence of locally synchronized and unsynchronized oscillators. In our simulation, we start with oscillators with random phases, {\it{i.e.}}  sampled from a normal distribution, with mean $0$ and variance $2\pi$. For $\alpha=1.43$ and $A>A_c$ where $A_c \approx 0.64$ 
for $\delta=0$, the system evolves to the chimera state, where the oscillators break into local patches of synchronized and unsynchronized oscillators. The chimera phase in the reciprocal version of the model was first observed by Kuramoto~\cite{kuramoto2002coexistence} and later analyzed by Abrams and Strogatz~\cite{Abrams2004}. Fig.~\ref{Fig7}$(a)$ shows a typical chimera state, with the $x$-axis denoting space, the $y$-axis indicating time, and the color bar denoting the phases of each oscillator. 

\subsubsection{Run-and-Chase}

For larger values of $\delta$ (typically for $\delta > 0.3$), oscillators no longer form static synchronized clusters. Instead, each oscillator tends to synchronize with one of its immediate neighbors while trying to go out of synchrony with the other (a similar state has been observed in Ref.~\cite{Multicluster2014}). The asymmetric local interaction (as given in Eq.~\ref{kernel}), which is controlled by $\delta$, results in the emergence of a linear spatial profile of phase that travels with a constant speed. As shown in Fig.~\ref{Fig7}(d), the phase field exhibits diagonal spatiotemporal stripes in the kymograph, and the slope of these patterns provides a measure of the propagation velocity. Based on the nomenclature in the non-reciprocity literature~\cite{Fruchart2021Nonreciprocal}, we call this the ``run-and-chase'' state. 

\subsubsection{Traveling Chimera}
For larger values of $A$ ($A>A_c$) and $\delta$, the coherent (synchronized) and incoherent (unsynchronized) regions no longer remain stationary in time but instead drift with a finite velocity; the parameter range in which this state is stable is shown in the phase diagram in Fig.~\ref{fig:8}. This dynamical phase is known as a traveling chimera state. In Fig.~\ref{Fig7}(b), the coherent domain can be seen moving across the system over time. As seen in this figure this phase can also coexist with a run-and-chase phase. The propagation velocity can be quantified from the slope of the trajectory of the synchronous region. 
To analyze the spatiotemporal phase patterns, we subtracted the average angular drift of the largest synchronized domain at each time step. This procedure removes the systematic angular drift and effectively produces a co-moving reference frame, allowing us to isolate the intrinsic dynamics of the chimera state.

The traveling behavior (as well as the run-and-chase dynamics) can emerge only when the left–right symmetry of the coupling is broken, {\it{i.e.}} when the interactions are non-symmetric.
Left–right symmetry can be broken by varying the shape of the interaction kernel and traveling chimera states have been reported in the Kuramoto model under such symmetry-breaking couplings~\cite{TCWeakTemporal, Smirnov2024}. It will be interesting to explore in the future. whether our choice of breaking the left-right symmetry is crucial for getting  the chimera II state (discussed in the next section).

\subsubsection{Chimera II}

\begin{figure}
    \centering
    \includegraphics[width=\linewidth]{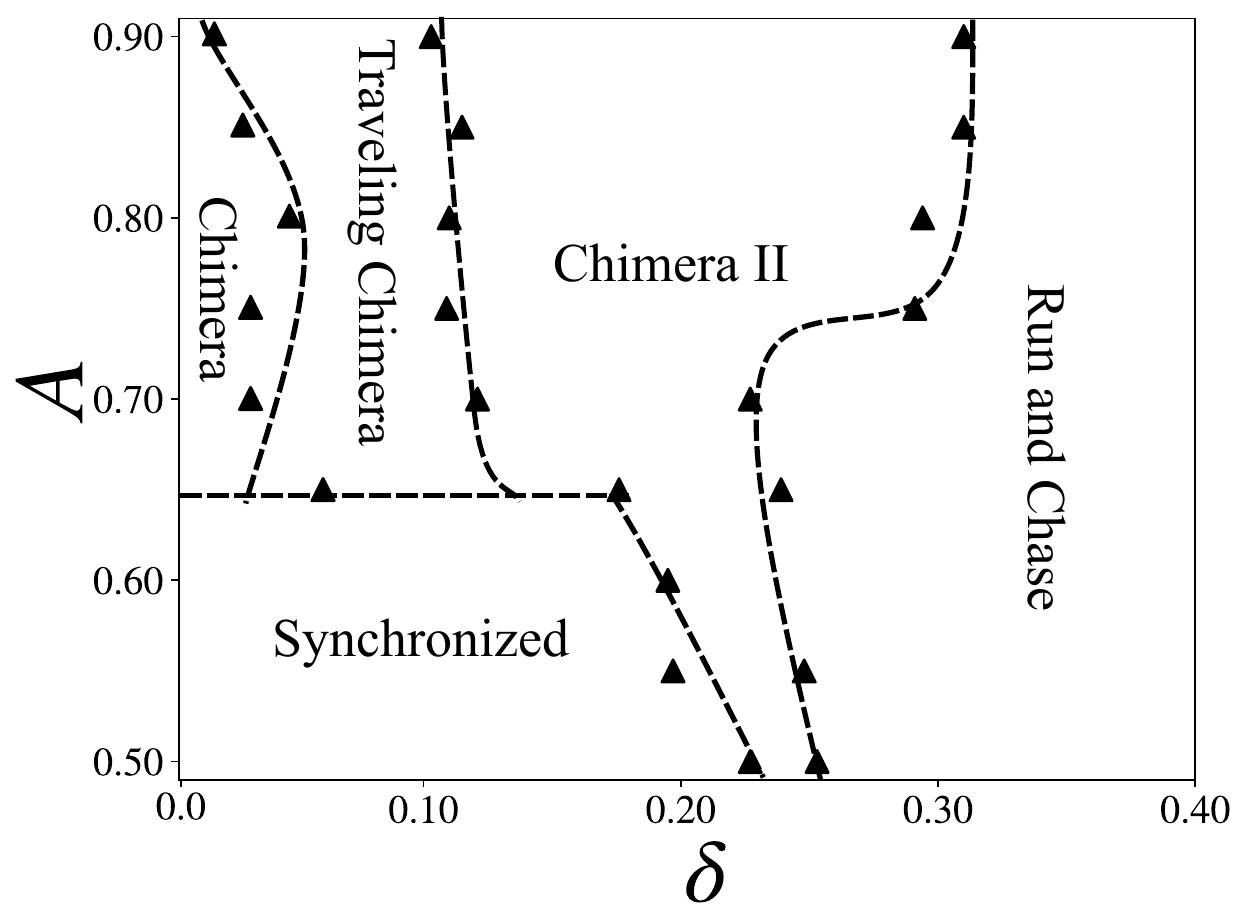}
    \caption{Phase diagram of the spatially extended model in the $(A,\delta)$ plane, showing the five observed phases -- synchronized, chimera, traveling chimera, chimera II 
    and run-and-chase --and their respective stability regions for $\alpha = 1.43$. The diagram was obtained through direct numerical simulations. The phase boundaries where the transition between two regimes occurs are marked with solid symbol. The dashed lines are a guide to eye for the phase boundaries.}
    \label{fig:8}
\end{figure}

When we increase $\delta$ (into the range  $\delta \gtrsim 0.2$ for $A<A_c$ and  $\delta \gtrsim 0.15$ for $A>A_c$) the population once again splits into two distinct groups. But, in contrast to the chimera state, we observe in this regime local patches of synchronized oscillators coexisting with regions where oscillators display run-and-chase dynamics; see Fig.~\ref{Fig7}(c) for the kymograph of such a state, where synchronized clusters are interspersed with domains exhibiting run-and-chase dynamics. We refer to this coexistence state as a type II chimera: see Fig.~\ref{fig:8} for the phase diagram indicating where this state can be found. 

\subsubsection{Phase Diagram}

The phases mentioned above are observed at different values of $A$ and $\delta$ for a particular $\alpha$. We have kept $\alpha$ fixed at $\alpha = 1.43$ for the one-dimensional model and then carried out numerical integration of Eq.~\ref{eq14} for different combinations of $A$ and $\delta$. For each parameter set, oscillator phases were initialized randomly 
, drawn from a normal distribution with mean $0$ and variance $2\pi$.  Each simulation was repeated ten times, and the observed phases are reported on the basis of more than 50\% appearance of the corresponding dynamical state. Fig. \ref{fig:8} summarizes all the phases obtained numerically -- as exemplified in Fig.~\ref{Fig7} -- into a phase diagram in the ($\delta,A$) plane. 

\section{Discussion}

In this paper, we demonstrated that the introduction of non-reciprocal interactions in coupled oscillator populations not only modifies the phase diagram of the system—shifting the transition points between different phases, but also gives rise to novel phases absent in the corresponding reciprocal model~\cite{Abrams2004,Abrams2008}.
To establish this, we first analyzed a model of two oscillator populations with inter- and intra-population couplings, extending the reciprocal case studied by Abrams et al.~\cite{Abrams2008} to include non-reciprocal inter-population interactions. This modification leads to qualitative changes in the phase diagram, confirmed analytically via the mean field theory and the Ott–Antonsen ansatz~\cite{Ott2008}, and produces a run-and-chase phase.

We then investigated a spatially extended version of the Kuramoto model, where non-reciprocity was introduced through asymmetric interactions breaking the right-left symmetry. This yielded additional nontrivial dynamical regimes, including a traveling chimera phase and a second form of chimera state (a coexistence of the synchronized and run \& chase state), alongside the pure run-and-chase behavior.
The phase diagram for the second model reveals that the system transitions from a fully synchronized state to a run and chase phase while increasing $\delta$ at a small $A<A_c$. This happens via a coexistence regime where synchronized oscillators and run-and-chase dynamics coexist; we refer to this mixed state as chimera II. 
For larger values of $\text{A}$, we observe a different progression: at small $\delta$, a chimera state still persists; however, it becomes spatially mobile and exhibits coherent drift over time, which we classify as a traveling chimera \cite{TCWeakTemporal, TCPendula}. When $\delta$ is further increased, the traveling chimera transitions back into the chimera-II state, which ultimately converges to the run-and-chase phase at even  higher $\delta$. A detailed bifurcation analysis of these transitions has not been carried out in the present work and is left for future investigation.  

Our results highlight the rich phenomenology introduced by non-reciprocal coupling and will help us better understand and control exotic phases such as chimera via non-reciprocity. Our study will motivate similar extensions to oscillator models in higher dimensions ({\it{e.g.}} using vision cone type interaction range~\cite{Bandini2025NRXY, Demian_Levis_Vision_Cone_XY}) and to systems beyond the Kuramoto class. Experimental realizations, for example, in artificial oscillator arrays with tunable asymmetric hydrodynamic couplings~\cite{BratoPRL}, could provide direct tests of our predictions. 

\section*{Acknowledgement}

S.M.C. acknowledges RRI for using the HPC facility for computational purpose. R.M. acknowledges support from the ANRF, India, through PMECRG (project ANRF/ECRG/2024/002036/PMS).

\bibliography{references}

\clearpage
\externaldocument{main}
\renewcommand{\theequation}{S\arabic{equation}}
\renewcommand{\thesection}{S\arabic{section}}
\renewcommand{\thesubsection}{S\arabic{section}.\arabic{subsection}}
\renewcommand{\thefigure}{S\arabic{figure}}
\renewcommand{\thetable}{S\arabic{table}}
\setcounter{section}{0}
\setcounter{figure}{0}
\setcounter{table}{0}
\setcounter{equation}{0}

\section*{Supplementary Information}
\section{Run and Chase for Two Population Model}

In this appendix, we derive the reduced mean-field dynamics governing the phase difference $\Theta$ and the collective mean phase $\Phi$ between two coupled oscillator populations.  
We consider two populations of identical phase oscillators, labeled $1$ and $2$, each consisting of $N$ oscillators. The phase dynamics are given by the Kuramoto equations:
\begin{align}
\dot{\theta}_i^1 &= \omega + \sum_{j=1}^N \left[\frac{K_{11}}{N}\sin(\theta_j^1 - \theta_i^1 - \alpha) + \frac{K_{12}}{N}\sin(\theta_j^2 - \theta_i^1 - \alpha)\right] \\
\dot{\theta}_k^2 &= \omega + \sum_{l=1}^N \left[\frac{K_{21}}{N}\sin(\theta_l^1 - \theta_k^2 - \alpha) + \frac{K_{22}}{N}\sin(\theta_l^2 - \theta_k^2 - \alpha)\right]
\end{align}

where $\omega$ is the intrinsic frequency and $\alpha$ is the phase-lag parameter. The coupling matrix is given by:
\begin{equation}
K = \begin{pmatrix}
\frac{1+A}{2} & \frac{1-A}{2} + \delta \\
\frac{1-A}{2} - \delta & \frac{1+A}{2}
\end{pmatrix}
\end{equation}

where $A$ is the disparity between the intra- and inter-population couplings, and $\delta$ is the non-reciprocity parameter. 

\subsubsection{Mean-field Dynamics}
For the fully synchronized populations, we may replace all the oscillators within a population by a single effective phase variable: $\theta_{i}^{1}\equiv \theta_{1}$ and $\theta_{k}^{2}\equiv \theta_{2}$. In this case, each population behaves as a single oscillator with a mean-field dynamics
\begin{align}
\dot{\theta}_1 &= \omega + K_{11}\sin(-\alpha) + K_{12}\sin(-\Theta - \alpha) \\
\dot{\theta}_2 &= \omega + K_{21}\sin(\Theta - \alpha) + K_{22}\sin(-\alpha)
\end{align}

We can now define the phase difference and phase sum as $\Theta = \theta_{2} - \theta_{1}$ and $\Phi =\frac{\theta_{1} + \theta_{2}}{2} $. Now we will derive the evolution equation for $\Theta$ and $\Phi$. 

\subsubsection{Phase Difference Equation}

Subtracting $\dot{\theta}_1$ from $\dot{\theta}_2$ we obtain:
\begin{equation}
\dot{\Theta} = -2\delta\sin{(\alpha)}\cos{(\Theta)} + (1-\text{A})\cos{(\alpha)}\sin{(\Theta)}
\end{equation}

\subsubsection{Steady State Phase Difference}
At steady state ($\dot{\Theta} = 0$):
\begin{align}
(1-A)\sin\Theta\cos\alpha &= 2\delta\cos\Theta\sin\alpha \\
\tan\Theta &= \frac{2\delta\sin\alpha}{(1-A)\cos\alpha}
\end{align}
\begin{equation}
    \Theta = \arctan\left(\frac{2\delta}{1-A}\tan\alpha\right)
    \label{theta}
\end{equation}

In the main text in Fig.\ref{Fig4}(c) we show the collapse of the expression.

\subsubsection{Phase Sum Dynamics and the Collective Frequency}
To compute the collective frequency, we consider the evolution of the phase sum, whose derivative is given by the following equation:

\begin{align}
\dot{\Phi} &= + K_{11}\sin(-\alpha) + K_{12}\sin(\Theta - \alpha) \\
&\quad + K_{21}\sin(-\Theta - \alpha) + K_{22}\sin(-\alpha)
\end{align}

We can set the $\omega = 0$. Also, in the steady state dynamics, if Eq.\ref{theta} is satisfied, then few lines of algebraic manipulation shows that, 

\begin{equation}
    \dot{\theta}_{1} = \dot{\theta}_{2} = \dot{\Phi}
\end{equation}
Hence, it shows the time evolution of the steady state phase velocity $\dot{\Phi}$ can be written down in terms of the Kuramoto equation for the two-population oscillators.

\begin{equation}
    \dot{\Phi} = \frac{1+A}{2} \sin{(-\alpha)} + (\frac{1-A}{2} + \delta)\sin(\Theta-\alpha)
\end{equation}

Rearranging this term results in the collapse shown in Fig.\ref{Fig4}(d),

\begin{equation}
    \frac{\dot{\Phi} + \frac{1+A}{2}\sin{\alpha}}{\frac{1-A}{2} + \delta}  = \sin{(\Theta - \alpha})
\end{equation}

\section{Non-reciprocity in $\alpha$}\label{appendix NR in alpha}

In the main text, we have explored in detail the effects of non-reciprocity in the inter-population interactions of a two-population oscillator model. One can also introduce non-reciprocity in the phase lag parameter $\alpha$. In that case the oscillators of population $1$, maintain a phase difference of $\alpha +\delta$ with oscillators of population $2$, and that of population $2$ tends to maintain a phase difference of $\alpha -\delta$. 

\begin{figure}
    \centering
\includegraphics[width=1\linewidth]{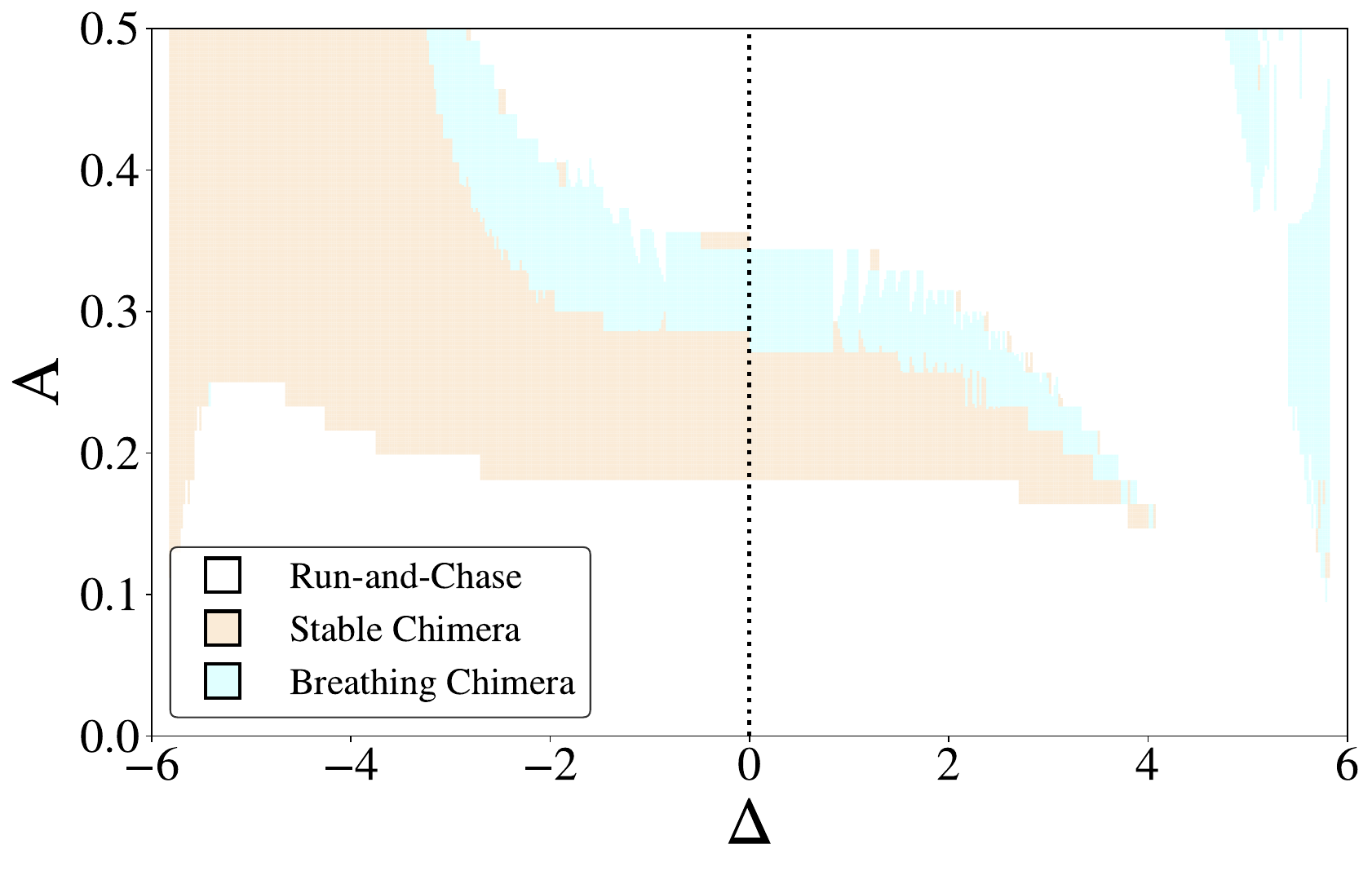}
    \caption{Phase diagram describing the stability of states in the $A-\delta$ plane for the non-reciprocity in the phase-lag parameter $\alpha$ where $\alpha = 1.47$. The $x$-axis is rescaled for visual ease with $\Delta = sgn(\delta)\ln(1+\frac{|\delta|}{\epsilon})$, where $\epsilon=0.003$ is the scaling parameter.}
    \label{fig:S1}
\end{figure}

We carried out numerical integration of Eq.~\ref{eq1} (main text) with non-reciprocity in $\alpha$, by making them $\alpha_{12} = \alpha +\delta$ for oscillators of population 1 and $\alpha_{21} = \alpha -\delta$ for oscillators of population 2 (here, we have kept the phase lag $\alpha$ to be the same, as the oscillators in both the populations are considered identical). The alpha matrix then becomes, 
\begin{eqnarray}
    \boldsymbol{\alpha} \equiv \begin{pmatrix}
        \alpha_{11} &  \alpha_{12} \\
        \alpha_{21} & \alpha_{22}
        
    \end{pmatrix} = \begin{pmatrix}
        \alpha &  \alpha + \delta \\
        \alpha - \delta & \alpha
        
    \end{pmatrix}
\end{eqnarray}
We report a significant change in the phase boundaries in the $A-\delta$ plane due to the non-reciprocity in phase lag parameter $\alpha$, which is shown in Fig.\ref{fig:S1}. 

\section{Phase Diagram for different $\alpha$}

In the main text, we demonstrated that introducing non-reciprocity $\delta$ in the inter-population couplings can significantly modify the regions of chimera states in the phase diagram (the $A-\delta$ plane). While the phase diagram discussed in the main text corresponds to a fixed phase-lag parameter $\alpha = 1.47$, the analysis can be readily extended to other values of $\alpha$ without any loss of generality. In this section, however, we specifically wanted to focus on those values of $\alpha$ for which the system does not exhibit chimera states under reciprocal coupling ($\delta = 0$).

\begin{figure}
    \centering
    \includegraphics[width=1\linewidth]{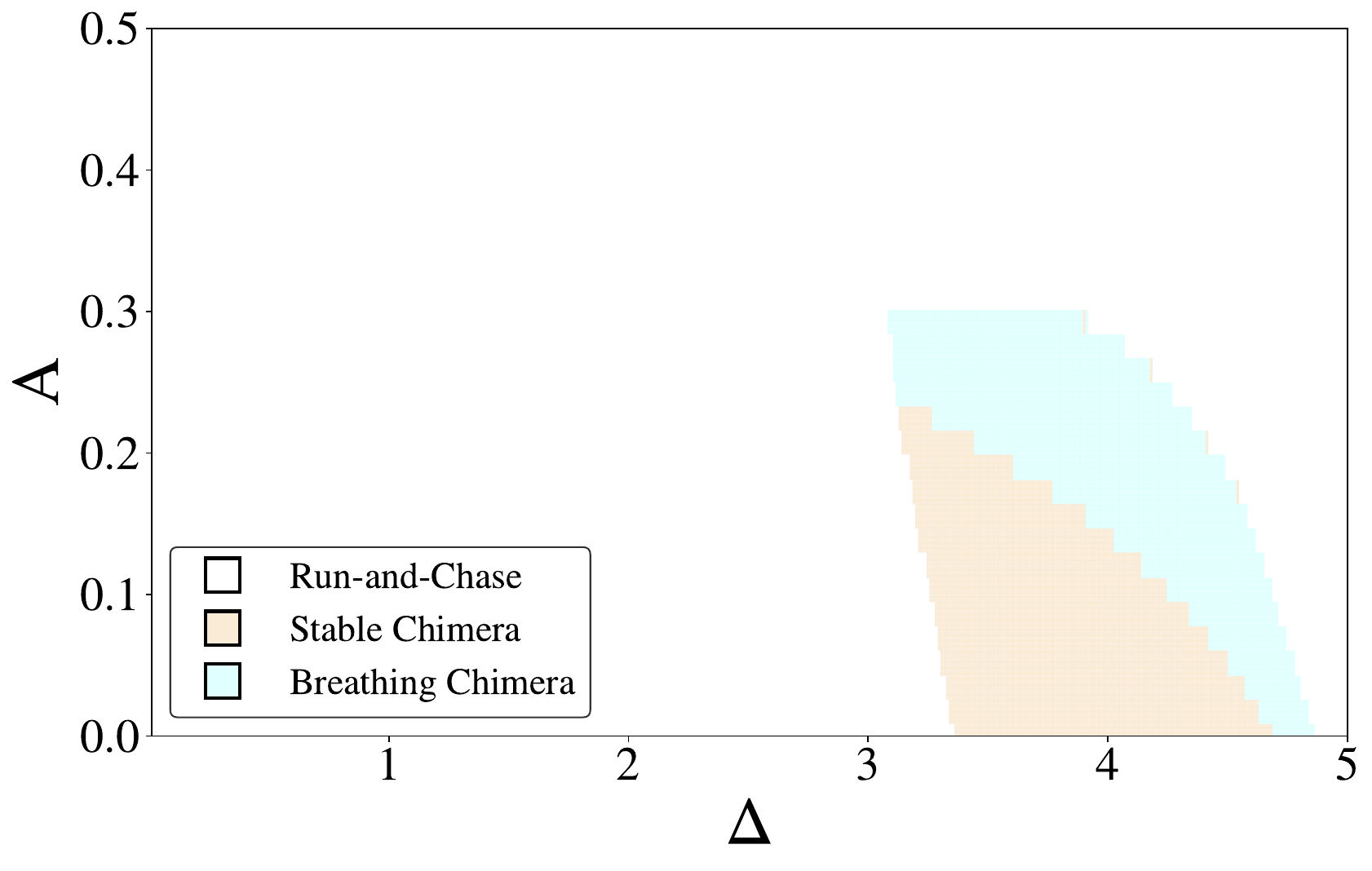}
    \caption{Phase diagram of the stability of states in the $A-\delta$ plane, constructed by running simulations where $\alpha=1.3$. The $x$-axis is rescaled for visual ease with $\Delta = sgn(\delta) \ln(1 + \frac{|\delta|}{\epsilon})$, where $\epsilon= 0.003$ is a scaling parameter.}
    \label{fig:S2}
\end{figure}

Numerical simulations reveal that by increasing $\delta$, basins of stability for the chimera states can emerge even in such cases. For instance, at $\alpha = 1.3$, no chimera was observed for $\delta = 0$, but as $\delta$ increases beyond a threshold, both breathing and stable chimeras appear in the phase diagram. This behavior is illustrated in Fig.~\ref{fig:S2}.

\section{Derivation of mean field amplitude equations}\label{appendix_meanfield_2d}

To get the 2D equation for the modulus and phase difference of the complex amplitude, Eq.~\ref{2variable_reduced_eqn}, we start from the continuity equation~\ref{continuity}
\begin{equation}
    \frac{\partial f^\sigma}{\partial t} + \frac{\partial}{\partial\theta} (f^\sigma v^\sigma) = 0
\end{equation}
To work in the continuum limit, we define complex order parameter~\cite{Abrams2008}:
\begin{equation}
    z_{\sigma}(t) = \sum_{\sigma'=1}^2 K_{\sigma\sigma'}\int e^{i\theta'}f^{\sigma'}(\theta', t) d\theta'
\end{equation}
Then, the velocity can be written in the form
\begin{equation}
    v^{\sigma}(\theta) = \omega + \frac{1}{2i}\left( z_{\sigma}e^{-i\alpha}e^{-i\theta} - z^*_{\sigma}e^{i\alpha}e^{i\theta} \right)
\end{equation}
Finally, for the solution of the Eq.~\ref{continuity}, using the Ott-Antonsen ansatz~\cite{Ott2008}
\begin{equation}
    f^\sigma(\theta,t) = \frac{1}{2\pi}\left[ 1+\left[ \sum_{n=1}^\infty (a_{\sigma}(t)e^{i\theta})^n +c.c \right] \right]
\end{equation}
it is possible to reduce the continuity equation into a single amplitude equation
\begin{equation}\label{amplitude_continuity_eqn}
    \dot{a}_\sigma+i \omega a_\sigma +\frac{1}{2} \left[a_\sigma^2 z_\sigma e^{-i\alpha} -z_\sigma^* e^{i\alpha}\right] = 0
\end{equation}
To close the equation, we input the ansatz in the expression for the order parameter, $z_\sigma$
\begin{equation}
    z_\sigma = \sum_{\sigma'} K_{\sigma\sigma'}a_{\sigma'}^*
\end{equation}
using orthogonality property. Inputting this into the amplitude equation,~\ref{amplitude_continuity_eqn},
\begin{eqnarray}
    &0&=\dot{a}_1 +i\omega a_1 +\frac{1}{2} a_1^2 (K_{11}a_1^*+K_{12}a_2^*) e^{-i\alpha}\cr
    &-& \frac{1}{2} (K_{11} a_1 + K_{12} a_2) e^{i \alpha}
\end{eqnarray}
and similarly for $a_2$ by switching labels. Now, rewriting the equation in polar coordinates, $a_\sigma = \rho_\sigma e^{-i\phi_\sigma}$
\begin{eqnarray}
    &\dot{\rho}_1& + \frac{1}{2} (\rho_1 ^2-1) \left[ \mu \rho_1\cos\alpha +(\nu+\alpha) \rho_2 \cos(\phi_2 -\phi_1 -\alpha) \right] = 0,\cr
    &\rho_1&\dot{\phi}_1 + \omega\rho_1 \cr
    &-&\frac{1+\rho_1^2}{2} [ \mu\rho_1 \sin\alpha +(\nu+\alpha) \rho_2 \sin(\phi_2 -\phi_1 + \alpha) ]= 0,\cr
    &\dot{\rho}_2& + \frac{1}{2} (\rho_2 ^2-1) \left[ \mu \rho_2\cos\alpha +(\nu-\alpha) \rho_2 \cos(\phi_1 -\phi_2 -\alpha) \right] = 0,\cr
    &\rho_2&\dot{\phi}_2 + \omega\rho_2 \cr
    &-&\frac{1+\rho_2^2}{2} [ \mu\rho_2 \sin\alpha +(\nu-\alpha) \rho_1 \sin(\phi_1 -\phi_2 + \alpha) ]= 0
\end{eqnarray}
For chimera state, oscillators of one population are synchronized. Following~\cite{Abrams2008}, we take this to be population 1 and put $\rho_1=1$
\begin{eqnarray}
    \dot{\phi}_1 &=& \omega - \mu \sin\alpha -(\nu+\alpha) r \sin(\Theta+\alpha)\cr
            \dot{r} &=& \frac{1-r^2}{2}\left[ \mu r\cos\alpha +(\nu-\alpha) \cos(\Theta-\alpha)  \right]\cr
            \dot{\phi}_2 &=& \omega -\frac{1+ r^2}{2r} \left[ \mu r\sin\alpha + (\nu-\alpha)\sin(\alpha - \Theta) \right]
\end{eqnarray}
using $\Theta=\phi_2-\phi_1$. Further, subtracting the phase equations, we get
\begin{eqnarray}
    \dot{r} &=& \frac{1-r^2}{2}(\mu r \cos\alpha+(\nu-\delta)\cos(\Theta-\alpha))\cr
    \dot{\Theta} &=& \frac{1+r^2}{2r}\left[ \mu r\sin\alpha -(\nu-\delta)\sin(\Theta-\alpha) \right]\cr
    &-&\mu \sin\alpha -(\nu+\delta) r\sin(\Theta+\alpha)
\end{eqnarray}

\section{Invariant under $\theta \to \theta- \omega t$}

Here, we provide a brief discussion on the invariance of the phase evolution equation of the two populations of Kuramoto oscillators under change from a stationary to a rotating frame of reference. The original equation is of the form \cite{Abrams2008}

\begin{equation}
    \frac{d \theta_{i}^{\sigma}}{dt} = \omega + \sum_{\sigma^{\prime} = 1}^{2} \frac{K_{\sigma\sigma^{\prime}}}{N_{\sigma^{\prime}}} \sum_{j=1}^{N_{\sigma^{\prime}}}\sin(\theta_{j}^{\sigma^{\prime}} - \theta_{i}^{\sigma} - \alpha).
    \label{Kuramoto eqn Original}
\end{equation}

Now, we can define the phases of the oscillators in the rotating frame of reference,  

\begin{equation}
    \tilde{\theta_{i}^{\sigma}} = \theta_{i}^{\sigma} - \omega t.
\end{equation}

Taking time derivatives
\begin{align}
\frac{d\tilde{\theta}_{i}^{\sigma}}{dt} 
    &= \frac{d\theta_{i}^{\sigma}}{dt} - \omega \\[4pt]
    &= \omega + \sum_{\sigma' = 1}^{2} \frac{K_{\sigma\sigma'}}{N_{\sigma'}} 
       \sum_{j=1}^{N_{\sigma'}} \sin(\theta_{j}^{\sigma'} - \theta_{i}^{\sigma} - \alpha) - \omega \\[4pt]
    &= \sum_{\sigma' = 1}^{2} \frac{K_{\sigma\sigma'}}{N_{\sigma'}} 
       \sum_{j=1}^{N_{\sigma'}} \sin(\theta_{j}^{\sigma'} - \theta_{i}^{\sigma} - \alpha) \\[10pt]
    &= \sum_{\sigma' = 1}^{2} \frac{K_{\sigma\sigma'}}{N_{\sigma'}} 
       \sum_{j=1}^{N_{\sigma'}} \sin\!\big( (\theta_{j}^{\sigma'} - \omega t) - (\theta_{i}^{\sigma} - \omega t) - \alpha \big) \\[10pt]
    &= \sum_{\sigma' = 1}^{2} \frac{K_{\sigma\sigma'}}{N_{\sigma'}} 
       \sum_{j=1}^{N_{\sigma'}} \sin\!\big( \tilde{\theta}_{j}^{\sigma'} - \tilde{\theta}_{i}^{\sigma} - \alpha \big)
\end{align}

This shows that setting $\omega = 0$ (which is equivalent to going to the rotating frame of reference) will not create any qualitative change in the system of oscillators.  

\begin{figure*}[]
    \centering\includegraphics[width=1\linewidth]{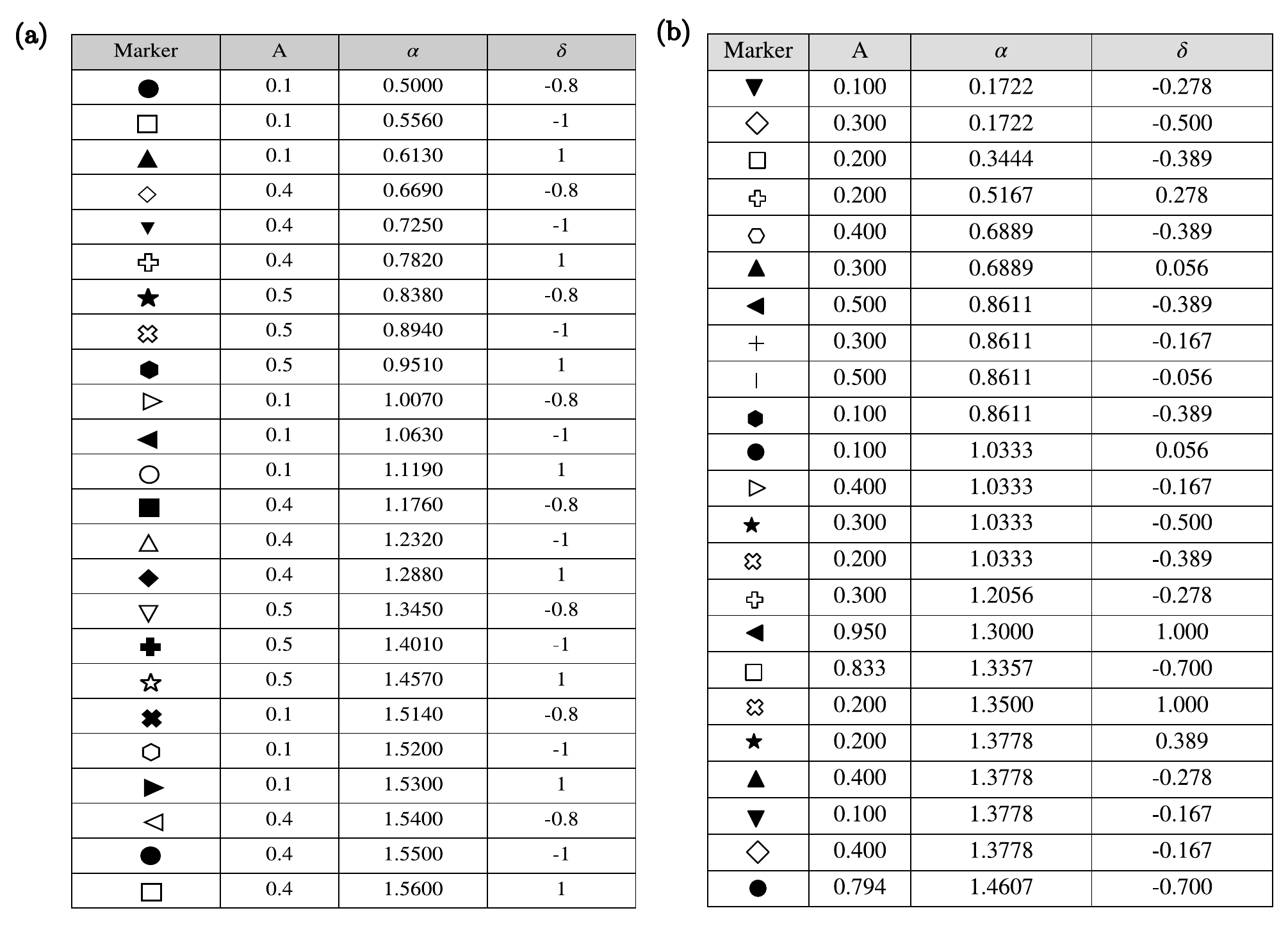}
    \caption{Different markers corresponding to different $A$, $\alpha$ and $\delta$ values for (a) Fig.\ref{Fig4}(c) and for (b) Fig.\ref{Fig4}(d), respectively.}
\end{figure*}

\section{Different Symbols for Fig. \ref{Fig4}}

\end{document}